# The Structure of Participation and Attention in Arabic-Language Hezbollah Discourse on X

Mohamed Soufan
Independent Researcher & Data Scientist
Antalya, Turkey
ORCID: https://orcid.org/0009-0004-1705-5574
Email: m@soufan.tech

## Abstract

Social media platforms play an increasingly important role in shaping political discussion and information flows. This study examines the structure of participation and attention in Arabic-language discourse about Hezbollah on X (formerly Twitter). Using a dataset of 15,767 tweets posted by 8,148 users between March 1 and March 8, 2026, the analysis investigates how engagement is distributed across participants and whether certain types of accounts play a disproportionate role in attracting attention.

The results reveal a highly unequal distribution of engagement. Although thousands of users participate in the conversation, the top 1% of users capture 61.5% of all engagement, while the top 10% capture 96.2%. At the same time, most content is produced by non-media users, who account for 89.6% of users and 79.9% of tweets in the dataset. Accounts labeled as media—identified through media-related keywords in account metadata—receive higher engagement per tweet on average (41.32 interactions) than non-media users (30.84 interactions) and are overrepresented among the most engaged accounts.

These findings indicate that while Hezbollah-related discourse on X appears broadly participatory in terms of posting activity, audience attention remains strongly concentrated among a small minority of highly visible accounts.

## 1. Introduction

Social media platforms have become central arenas for political discussion and information dissemination [1,2]. Platforms such as X (formerly Twitter) enable large numbers of users to participate in conversations about political actors, conflicts, and policy issues in real time. As a result, online discourse surrounding politically significant organizations often involves a diverse mix of participants, including ordinary users, journalists, activists, and media organizations.

Hezbollah is one of the most prominent and widely discussed political and military actors in the Middle East. Discussions about Hezbollah frequently occur across Arabic-language social media, where users debate political developments, share news coverage, and comment on regional events. Despite the apparent openness of these platforms, an important question remains: who actually captures attention within such conversations? While many users may participate in posting content, the distribution of audience engagement may be far more uneven. Previous research has shown that attention and influence on social media platforms are often highly concentrated among a relatively small number of users [3,4]. Recent work has also examined how linguistic features influence engagement in Arabic-language social media discourse [5].

This study examines the structure of participation and attention in Arabic-language discourse about Hezbollah on X. Using a dataset of 15,767 tweets posted by 8,148 users between March 1 and March 8, 2026, the analysis investigates how engagement is distributed across participants and whether certain types of accounts play a disproportionate role in attracting attention. Users were classified into two operational categories—media-labeled and non-media users—based on media-related keywords in account metadata, allowing comparison of posting activity and engagement patterns across these groups.

The findings reveal a pronounced concentration of attention within the conversation. Although the majority of tweets are produced by non-media users, engagement is highly concentrated among a small minority of accounts. In particular, the top 1% of users capture 61.5% of all engagement, while the top 10% capture 96.2%. In addition, accounts classified as media are disproportionately represented among the most engaged users and receive higher engagement per tweet on average. Together, these results suggest that while participation in Hezbollah-related discourse on X is broadly distributed, visibility and audience attention remain strongly concentrated among a small group of highly visible accounts.

## 2. Methodology

### 2.1 Data Collection

Data were collected from X (formerly Twitter) using automated queries executed through the Apify cloud platform. The collection process retrieved tweets returned by X search queries together with associated tweet metadata and account information.

Tweets were retrieved using Arabic-language search queries containing the term "حزب الله" and related spelling variations referring to Hezbollah. Each query included the lang:ar filter to restrict results to Arabic-language tweets and the -filter:retweets operator to exclude retweets. Because the retweet filter was applied directly within the search query, retweets were not collected at the data retrieval stage rather than being removed during later preprocessing.

Data collection was conducted through multiple scraping runs between March 1 and March 8, 2026. Because X search results return only a partial subset of matching posts, partially overlapping runs were used to improve coverage within the study window. The final dataset was restricted to tweets posted during the period March 1–8, 2026.

For each tweet, the collection process retrieved the tweet text, tweet ID, creation timestamp, engagement counts, and basic account metadata, including username, display name, profile bio, follower count, and verification status.

## 2.2 Dataset Construction and Preprocessing

Data were collected through multiple scraping runs covering partially overlapping time windows within the study period. Overlapping runs were used because X search results return only a limited subset of matching tweets, and repeated queries can retrieve additional posts that may not appear in a single search.

After collection, the outputs from all scraping runs were merged into a single dataset. Deduplication was performed using tweet IDs, ensuring that each tweet appeared only once in the final dataset even if it was retrieved in multiple scraping runs or through different search queries.

Basic preprocessing was then applied to remove non-human or automated accounts. In particular, tweets posted by the account `Grok`, X's built-in AI assistant, were removed because they represent automated responses rather than independent user participation in political discourse.

**Table 1 Summary statistics of the dataset by user type**

| Metric | Media | Non-media | Total |
| --- | --- | --- | --- |
| Users | 844 | 7,304 | 8,148 |
| Tweets | 3,164 (20.1%) | 12,603 (79.9%) | 15,767 |
| Avg. tweets/user | 3.75 | 1.73 | — |
| Total engagement | 130,737 | 388,725 | 519,462 |
| Avg. engagement/tweet | 41.32 | 30.84 | — |

Note: Engagement = likes + reposts + replies.

After deduplication and preprocessing, the final dataset contained 15,767 tweets posted by 8,148 unique users between March 1 and March 8, 2026.

## 2.3 User Classification

Users were classified into two categories: media-labeled accounts and non-media users. Classification was performed using a keyword-based approach applied to account metadata, including the username, display name, and profile bio.

Accounts were labeled as media if these fields contained media-related keywords indicating association with journalism or media activity (e.g., channel, newspaper, broadcaster, agency, or journalist-related descriptors in Arabic or English).

Because keyword-based approaches may occasionally misclassify accounts, a limited number of manual overrides were applied. Accounts confirmed as institutional media outlets were assigned to a FORCE_MEDIA list, while accounts that matched media-related keywords but were clearly not institutional media accounts (such as parody accounts) were assigned to a FORCE_NON_MEDIA list.

Following this procedure, the final dataset contains 844 media-labeled users (10.4%) and 7,304 non-media users (89.6%).

## 2.4 Engagement Metric

Engagement was operationalized as the sum of likes, reposts, and replies received by a tweet. For each tweet, engagement was calculated as:

engagement = likeCount + retweetCount + replyCount

This metric captures the primary forms of user interaction on the platform and is commonly used to measure audience response to social media content.

Quote tweets were not included in the engagement metric. Although the dataset contains a `quoteCount` field, quote interactions occur on the quoting tweet rather than the original tweet, which introduces ambiguity about attribution. To maintain a consistent and comparable measure of interaction, the engagement metric was therefore restricted to likes, reposts, and replies.

Engagement values were aggregated both at the tweet level and the user level. Tweet-level engagement was used to calculate average engagement per tweet, while user-level totals were used to examine how engagement is distributed across participants in the conversation.

## 2.5 Analytical Approach

The analysis focuses on the relationship between participation (who produces content) and attention (who receives engagement) in Hezbollah-related discourse on X.

First, participation patterns were examined by comparing the number of users and tweets associated with media-labeled accounts and non-media users. Second, engagement patterns were analyzed by calculating the average engagement per tweet and the share of total engagement for each group.

To evaluate how attention is distributed across users, the study also measures engagement concentration. Users were ranked by their total engagement, and the cumulative share of engagement captured by the top 1%, 5%, and 10% of users was calculated. This approach allows the analysis to assess the extent to which a small subset of accounts dominates audience attention within the conversation.

Together, these descriptive analyses provide a structural overview of participation and engagement dynamics in the dataset without relying on statistical modeling.

## 3. Results

### 3.1 Engagement Concentration

Engagement in Hezbollah-related discourse on X is highly concentrated among a small fraction of users.
Although the dataset contains 8,148 unique users, the majority of attention is captured by a very small group of accounts.

The top 1% of users (81 accounts) capture 61.5% of all engagement in the dataset.
This concentration becomes even more pronounced at higher thresholds: the top 5% of users (407 accounts) capture 90.6% of total engagement, while the top 10% (815 accounts) capture 96.2% (Figure 1).

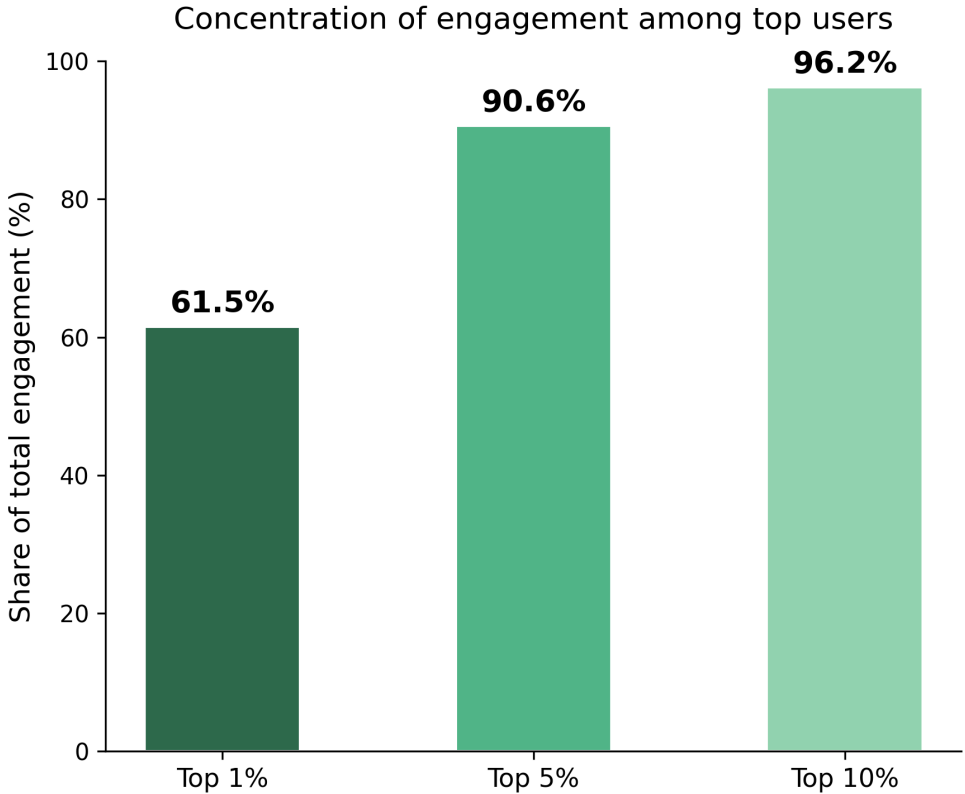

*Data: Arabic Hezbollah-related tweets on X, Mar 1–8, 2026 (N = 15,767 tweets; 8,148 users).*

Figure 1 Concentration of engagement among top users. The top 1%, 5%, and 10% of users capture 61.5%, 90.6%, and 96.2% of total engagement.

These results indicate that while thousands of users participate in the conversation, attention on the platform is overwhelmingly focused on a small minority of highly visible accounts.

### 3.2 Participation Structure

Although engagement is highly concentrated, participation in Hezbollah-related discourse on X is broadly distributed across users.

The dataset contains 8,148 unique users, of which 7,304 (89.6%) are non-media users and 844 (10.4%) are media accounts (Figure 2).

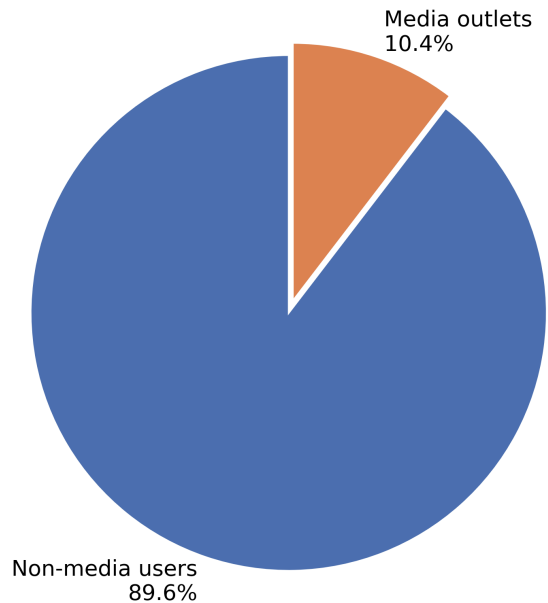

*Data: Arabic Hezbollah-related tweets on X, Mar 1–8, 2026 (N = 15,767 tweets; 8,148 users).*

Figure 2 User composition in Hezbollah-related discourse on X.

Non-media users account for 89.6% of participants, while media accounts represent 10.4% of users in the dataset.

Media accounts include institutional news organizations and verified media outlets, while non-media users consist of individuals, commentators, activists, and other accounts.

In terms of content production, non-media users generate the majority of posts. Out of 15,767 tweets in the dataset, 12,603 tweets (79.9%) were produced by non-media users, compared with 3,164 tweets (20.1%) from media accounts (Figure 3).

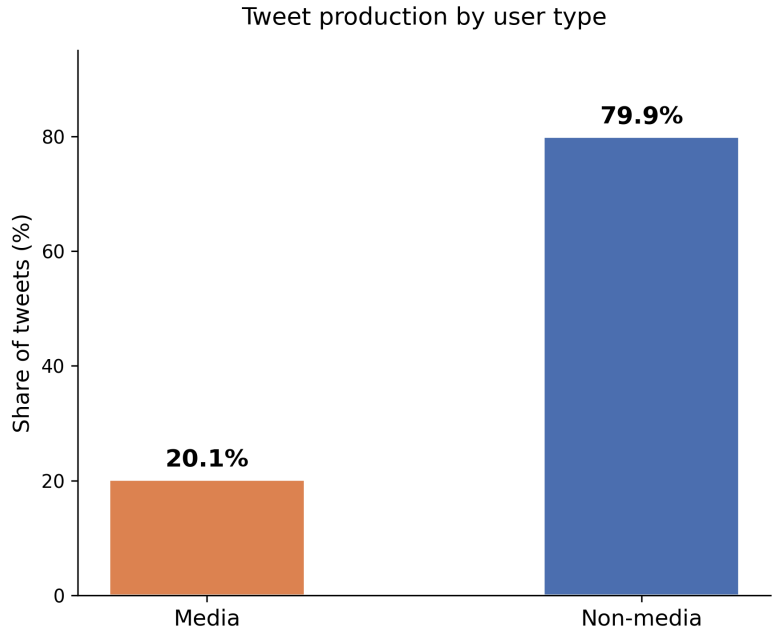

*Data: Arabic Hezbollah-related tweets on X, Mar 1–8, 2026 (N = 15,767 tweets; 8,148 users).*

Figure 3 Tweet production by user type. Non-media users generate 79.9% of tweets in the dataset, while media accounts produce 20.1%.

These results indicate that discussion about Hezbollah on X is largely driven by ordinary participants in terms of posting activity, even though engagement on the platform is concentrated among a much smaller set of highly visible accounts.

## 3.3 Media Attention Advantage

Although media accounts produce a smaller share of tweets, their posts attract substantially higher engagement on average.

Across the dataset, tweets from media accounts receive 41.32 interactions per tweet on average, compared with 30.84 interactions per tweet for non-media users (Figure 4).

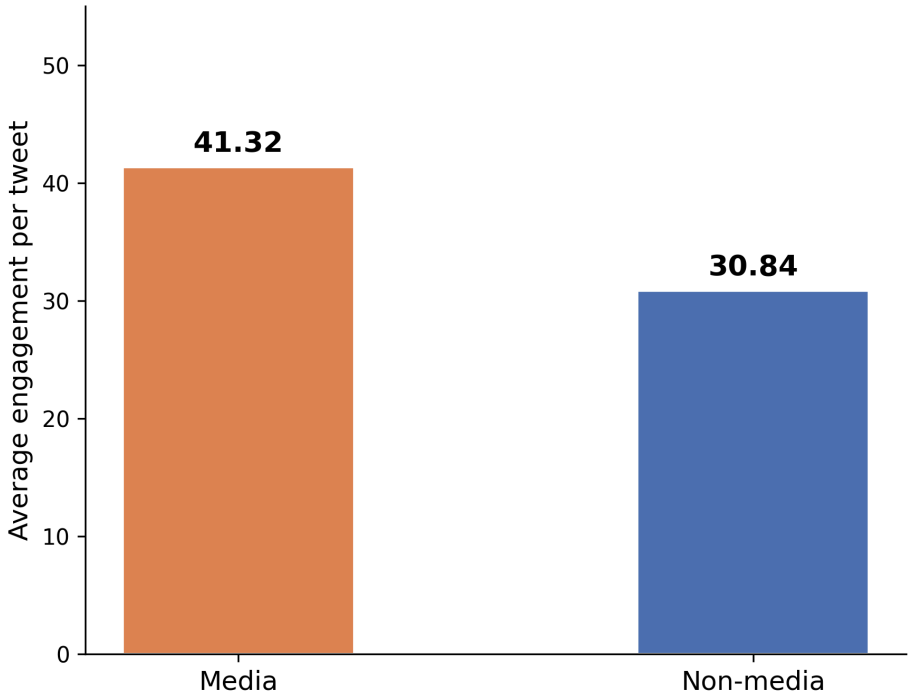

*Data: Arabic Hezbollah-related tweets on X, Mar 1–8, 2026 (N = 15,767 tweets; 8,148 users).*

Figure 4 Average engagement per tweet by user type. Tweets from media accounts receive 41.32 interactions per tweet on average, compared with 30.84 interactions for non-media users.

This indicates that media posts generate roughly 34% more engagement per tweet than those published by non-media accounts.

Despite this advantage, most engagement in the dataset is still generated by non-media users overall. Non-media accounts capture 74.8% of total engagement, while media accounts account for 25.2% (Figure 5).

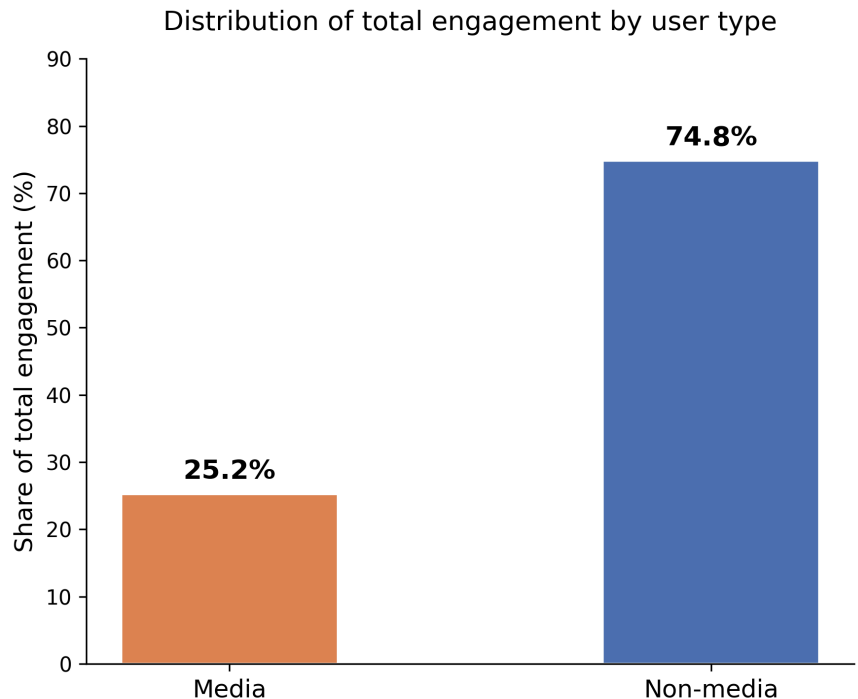

*Data: Arabic Hezbollah-related tweets on X, Mar 1–8, 2026 (N = 15,767 tweets; 8,148 users).*

Figure 5 Distribution of total engagement by user type. Non-media users account for 74.8% of total engagement, while media accounts capture 25.2%.

This pattern reflects the much higher volume of tweets produced by non-media participants.

Taken together, these findings suggest that media outlets retain a structural advantage in attracting attention on the platform, even though the majority of content is produced by non-media users.

### 3.4 Composition of the Most Engaged Users

To better understand the accounts that dominate engagement, we examine the composition of the top 1% of users ranked by total engagement.

This group consists of 81 users, who collectively capture 61.5% of all engagement in the dataset (see Section 3.1). Among these accounts, 24 (29.6%) are media outlets, while 57 (70.4%) are non-media users.

Although media accounts represent only 10.4% of all users in the dataset, they account for nearly 30% of the most engaged users, indicating that institutional media accounts are substantially overrepresented among highly visible accounts.

Manual inspection of these highly engaged users shows a heterogeneous mix of account types. The group includes institutional media outlets, journalists, political commentators, activists, official spokesperson accounts, and several large-following parody accounts. Most of these accounts also possess large follower bases, with the majority exceeding 10,000 followers and more than half exceeding 100,000 followers.

These observations suggest that the accounts capturing the majority of attention in Hezbollah-related discourse on X consist not only of institutional media organizations but also a broader set of influential public figures and highly followed online personalities.

## 4. Discussion

The results reveal a clear disparity between participation and attention in Hezbollah-related discourse on X. While the conversation involves thousands of users, engagement is overwhelmingly concentrated among a small minority of accounts. In this dataset of 8,148 users, the top 1% capture 61.5% of all engagement, while the top 10% capture 96.2%. This pattern indicates that although the conversation appears broadly participatory in terms of posting activity, visibility and audience attention are highly concentrated among a limited set of highly visible accounts. Similar patterns of highly unequal attention and influence have been documented in previous studies of online networks and social media platforms [3,4].

Media-labeled accounts also play a notable role within this attention structure. Although users classified as media represent only 10.4% of all users, they account for 29.6% of the most engaged accounts in the top 1%. In addition, tweets from media-labeled accounts receive higher engagement per tweet on average (41.32 interactions per tweet) compared with 30.84 interactions for non-media users. These patterns suggest that accounts associated with media-related identities—whether institutional outlets or individuals presenting themselves in media roles—tend to attract greater visibility within the conversation.

Inspection of the most engaged accounts suggests that influence within the conversation is distributed across a heterogeneous set of actors. The highly engaged group includes institutional media outlets, journalists, political commentators, activists, official spokesperson accounts, and several high-following parody or satirical accounts. Most of these accounts also have substantial follower bases, with the majority exceeding 10,000 followers and more than half exceeding 100,000 followers, indicating that existing audience reach likely plays an important role in shaping attention dynamics on the platform.

Finally, the classification used in this study should be interpreted as an operational proxy rather than a strict institutional distinction. Accounts were labeled as media based on the presence of media-related keywords in the username, display name, or profile bio, supplemented by a small number of manual overrides for clearly identifiable outlets. Consequently, some individual journalists whose profiles contained media-related descriptors were classified as media, while others were classified as non-media. The media category should therefore be understood as capturing accounts presenting themselves as media-related actors rather than representing a precise separation between institutional outlets and individual users.

## 5. Conclusion

This study examined the structure of participation and attention in Arabic-language discourse about Hezbollah on X using a dataset of 15,767 tweets posted by 8,148 users between March 1 and March 8, 2026. The findings reveal a clear distinction between broad participation and concentrated audience attention.

While the conversation involves thousands of users and is largely driven by non-media participants in terms of posting activity, engagement is highly concentrated among a small minority of accounts. In particular, the top 1% of users capture 61.5% of total engagement, while the top 10% capture 96.2%, indicating that visibility within the conversation is dominated by a limited set of highly visible users.

Accounts labeled as media are also disproportionately represented among the most engaged users and receive higher engagement per tweet on average than non-media users. Together, these patterns suggest that although Hezbollah-related discourse on X appears broadly participatory, audience attention remains strongly concentrated among a small group of influential accounts, including a notable share of media-labeled users.

Overall, the results illustrate how online political discussions can involve large numbers of participants while still exhibiting highly unequal distributions of visibility and engagement.

## Data and Code Availability

Due to the terms of service of X (formerly Twitter), the full tweet content collected for this study cannot be publicly redistributed. However, the analysis code used for data processing and statistical analysis will be made available through a public repository. Tweet identifiers and derived aggregate statistics may be shared upon reasonable request to support reproducibility of the results.

## References

[1] Hermida, A. (2010).
Twittering the news: The emergence of ambient journalism.
Journalism Practice, 4(3), 297–308.
https://doi.org/10.1080/17512781003640703

[2] Bruns, A., & Burgess, J. (2015).
Twitter hashtags from ad hoc to calculated publics.
In N. Rambukkana (Ed.), Hashtag Publics: The Power and Politics of Discursive Networks (pp. 13–28). Peter Lang.

[3] Barabási, A.-L. (2005).
The origin of bursts and heavy tails in human dynamics.
Nature, 435(7039), 207–211.
https://doi.org/10.1038/nature03459

[4] Bakshy, E., Hofman, J. M., Mason, W. A., & Watts, D. J. (2011).
Everyone's an influencer: Quantifying influence on Twitter.
Proceedings of the Fourth ACM International Conference on Web Search and Data Mining (WSDM '11) (pp. 65–74).

[5] Soufan, M. (2026).
Linguistic Uncertainty and Engagement in Arabic-Language X (formerly Twitter) Discourse.
arXiv:2603.00082.
https://arxiv.org/abs/2603.00082